\documentclass[preprint,aps,prl,showpacs,superscriptaddress]{revtex4}
\usepackage{graphicx}
\usepackage{bm}
\usepackage{amssymb}
\begin{document}
\title{Molecular Motor of Double-Walled Carbon Nanotube Driven by Temperature Variation}
\author{Zhan-chun \surname{Tu}}
\email[Email address: ] {tzc@itp.ac.cn} \affiliation{Institute of
Theoretical Physics,
 The Chinese Academy of Sciences,
 P.O.Box 2735 Beijing 100080, China}
 \affiliation{Graduate School,
 The Chinese Academy of Sciences, Beijing, China}
\author{Zhong-can \surname{Ou-Yang}}
\affiliation{Institute of Theoretical Physics,
 The Chinese Academy of Sciences,
 P.O.Box 2735 Beijing 100080, China}
\affiliation{Center for Advanced Study,
 Tsinghua University, Beijing 100084, China}
\begin{abstract}
An elegant formula for coordinates of carbon atoms in a unit cell
of a single-walled nanotube (SWNT) is presented and a new
molecular motor of double-walled carbon nanotube whose inner tube
is a long (8,4) SWNT and outer tube a short (14,8) SWNT is
constructed. The interaction between inner an outer tubes is
analytically derived by summing the Lennard-Jones potentials
between atoms in inner and outer tubes. It is proved that the
molecular motor in a thermal bath exhibits a directional motion
with the temperature variation of the bath.
\end{abstract}
\pacs{05.40.-a, 61.46.+w, 85.35.Kt} \maketitle
It is well known that a pollen in water exhibits Brownian motion.
The forces on the pollen stem from two components \cite{Astumian}:
A fluctuating force that averages to zero over time, and a viscous
force that slows the motion. These two kinds of forces are related
by temperature so the fluctuation is often called thermal noise.
The second law of thermodynamics suggests that biased Brownian
motion requires two conditions \cite{Reimann}: (1) breaking of
thermal equilibrium and (2) breaking of spatial inversion
symmetry. In order to illustrate the second law of thermodynamics
can not be violated, Feynman devised an imaginary ratchet system
with a pawl in his famous lecture \cite{Feynman}.

To explain the matter transport in biological systems, the concept
of molecular motor is introduced \cite{Yanagida,Kreis}. From the
statistical viewpoint, many models \cite{Reimann} of molecular
motor were put forward, such as on-off ratchets \cite{Bug,Ajdari},
fluctuating potential ratchets \cite{Astumian,Astumian2},
fluctuating force ratchets \cite{Magnasco,Luczka}, temperature
ratchets \cite{Reimann2,Li,Bao} and so on. All those models
satisfy the two conditions required by the second law of
thermodynamics.

With the development of nanotechnology, especially the discovery
of carbon nanotubes \cite{Iijima}, people are putting their dream
of manufacturing nanodevice \cite{Drexler} into practice. Here a
natural question arises: Can we construct molecular motors from
carbon nanotubes? In this Letter, we will prove that it is
possible to construct a molecular motor from a double-walled
carbon nanotube (DWNT). The molecular motor in a thermal bath
exhibits a directional motion with the temperature variation of
the bath.

\begin{figure}[htp!]
\includegraphics[width=7cm]{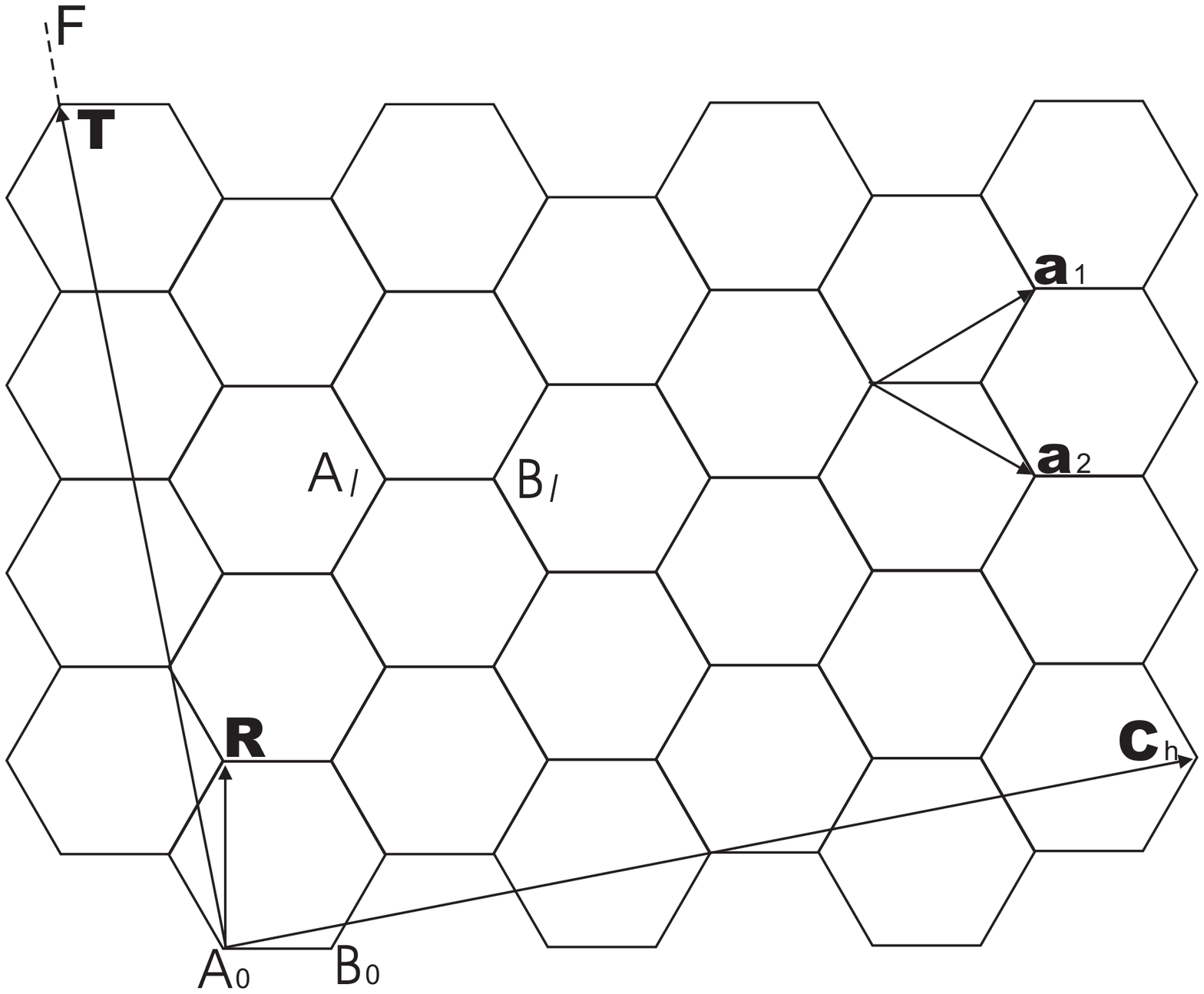}
\caption{\label{fig1}The unrolled honeycomb lattice of a SWNT. By
rolling up the sheet such that the point $A_0$ coincides with the
point corresponding to chiral vector ${\bf C}_h$, a nanotube is
formed. Vectors ${\bf a}_{1}$ and ${\bf a}_{2}$ are real space
unit vectors of the hexagonal lattice. The translational vector
${\bf T}$ is perpendicular to ${\bf C}_h$ and runs in the
direction of tube axis. The vector ${\bf R}$ is the symmetry
vector. $A_0$, $B_0$ and $A_l, B_l (l=1,2,\cdots,N)$ are used to
denote the sites of carbon atoms.}
\end{figure}

A DWNT consists of two single-walled carbon nanotubes (SWNT's)
with a common axis. The layer distance between the two tubes is
about $3.4$ \AA. A SWNT without two end caps can be constructed by
wrapping up a single sheet of graphite such that two equivalent
sites of hexagonal lattice coincide \cite{Saito}. To describe the
SWNT, some characteristic vectors need to be introduced. As shown
in Fig.\ref{fig1}, the chiral vector ${\bf C}_{h}$, which defines
the relative location of two carbon atoms, is specified by a pair
of integers $(n_1, n_2)$ which is called the index of the SWNT. We
have ${\bf C}_{h}=n_1{\bf a}_{1}+n_2{\bf a}_{2}$ with ${\bf
a}_{1}$ and ${\bf a}_{2}$ being two unit vectors of graphite. The
translational vector ${\bf T}$ corresponds to the first lattice
point of 2D graphitic sheet through which the line normal to the
chiral vector ${\bf C}_{h}$ passes. The unit cell of the SWNT is
the rectangle defined by vectors ${\bf C}_{h}$ and ${\bf T}$,
while vectors ${\bf a}_{1}$ and ${\bf a}_{2}$ define the area of
the unit cell of 2D graphite. The number $N$ of hexagons per unit
cell of the SWNT is obtained as a function of $n_1$ and $n_2$ as
$N=2(n_1^2+n_2^2+n_1n_2)/d_R$, where $d_R$ is the greatest common
divisor of ($2n_2+n_1$) and ($2n_1+n_2$). There are $2N$ carbon
atoms in each unit cell of the SWNT because each hexagon contains
two atoms. To denote the $2N$ atoms, we use a symmetry vector
${\bf R}$ to generate coordinates of carbon atoms in the nanotube;
it is defined as the site vector having the smallest component in
the direction of ${\bf C}_h$. From a geometric standpoint, vector
${\bf R}$ consists of a rotation around the nanotube axis by an
angle $\Psi=2\pi/N$ combined with a translation $\tau$ in the
direction of ${\bf T}$; therefore, ${\bf R}$ can be denoted by
${\bf R}=(\Psi|\tau)$. Using the symmetry vector ${\bf R}$, we can
divide the $2N$ carbon atoms in the unit cell of SWNT into two
classes: one includes $N$ atoms whose site vectors satisfy
\cite{tzc}
\begin{equation}\label{sitea}
{\bf A}_l=l{\bf R}-[l{\bf R}\cdot{\bf T}/{\bf T}^2]{\bf T} \quad
(l=0,1,2,\cdots,N-1),\end{equation} another includes the remainder
$N$ atoms whose site vectors satisfy
\begin{eqnarray}\label{siteb} &&{\bf B}_l=l{\bf R}+{\bf
B}_0-[\frac{(l{\bf R}+{\bf B}_0)\cdot{\bf T}}{{\bf T}^2}]{\bf
T}\nonumber
\\ &&-[\frac{(l{\bf R}+{\bf B}_0)\cdot{\bf C}_h}{{\bf C}_h^2}]{\bf C}_h
\quad (l=0,1,2,\cdots,N-1),\end{eqnarray}
 where {\bf B}$_0$
represents one of the nearest-neighboring atoms of {\bf A}$_0$. In
and only in the above two equations, $[\cdots]$ denotes the
Gaussian function, e.g., $[5.3]=5$.

\begin{figure}[htp!]
\includegraphics[width=7cm]{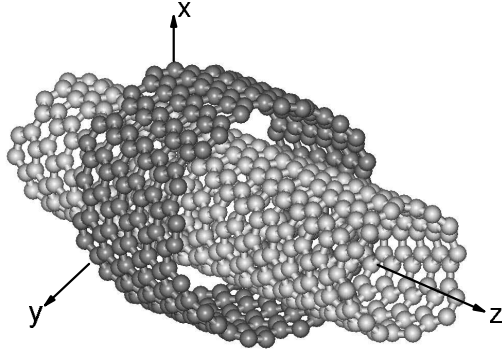}\caption{\label{fig2} A double-walled carbon nanotube
with the inner tube's index being (8, 4) and the outer tube's
index being (14, 8). $z$-axis is the tube axis, $x$-axis
perpendicular to $z$ passes through  carbon atom $A_0$ (see
Fig.\ref{fig1}) of the inner tube and $y$-axis is perpendicular to
the $xz$-plane. There is no obviously relative motion along radial
direction between the inner and the outer tubes at low
temperature. If we forbid the motion of the outer tube in the
direction of $z$-axis, only the rotation of the outer tube around
the inner tube is permitted.}
\end{figure}

We construct the molecular motor of double-walled carbon nanotube
as shown in Fig.\ref{fig2}. The inner tube's index is (8, 4) with
a length long enough to be regarded as infinite. The outer tube is
set to be a (14, 8) tube with just a single unit cell
\cite{remark1}. Obviously, they are both chiral nanotubes and
their layer distance is about 3.4 \AA. If we prohibit the motion
of the outer tube in the direction of nanotube axis, it will be
proved that this system in a thermal bath exhibits a directional
rotation when the temperature of the system varies with time. Thus
it could serve as a thermal ratchet.

To see this, we first select an orthogonal coordinate system shown
in Fig.\ref{fig2} whose $z$-axis is the tube axis and $x$-axis
passes through carbon atom $A_0$ of the inner tube. We fix the
inner tube and forbid the $z$-directional motion of the outer
tube, that is, carbon atom $A_0$ of outer tube is always in the
$xy$-plane. We denote the angle rotated by the outer tube around
the inner tube to be $\theta$.

\begin{figure}[htp!]
\includegraphics[width=7cm]{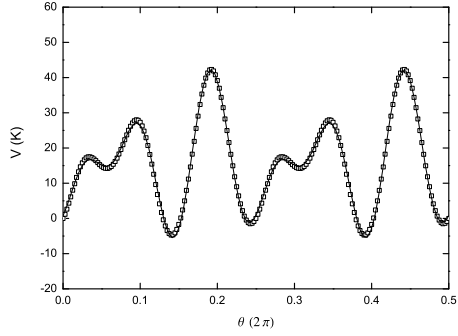}
\caption{\label{fig3} The potentials $V(\theta)$ between the outer
and the inner tubes with the outer tube rotating around the inner
tube. $\theta$ is the rotating angle. Here we have set $V(0)=0$.
The squares are the numerical results which can be well fitted by
$V(\theta)=15.7-0.6\cos
4\theta+-2.2\sin4\theta-12.7\cos8\theta-6\sin8\theta-1.7\cos12\theta+10.8\sin
12\theta$ (solid curve).}
\end{figure}

We take the interaction between atoms in the outer tube and the
inner tube as the Lennard-Jones potential
$u(r_{ij})=4\epsilon[(\sigma/r_{ij})^{12}-(\sigma/r_{ij})^6]$,
where $r_{ij}$ is the distance between atom $i$ in the inner tube
and atom $j$ in the outer tube, $\epsilon=28$ K, and
$\sigma=3.4$\AA \cite{hir}. We calculate the potential $V(\theta)$
\cite{tzc} when the outer tube rotates around the inner tube with
angle $\theta$ and plot it in Fig.\ref{fig3}. We find that
$V(\theta)$ is periodic (with period $\pi/2$) and violates the
spatial inversion symmetry. Thus the second condition to make a
molecular motor is satisfied. Up to now we can explain why we
select (8,4) tube and (14,8) tube. The following three criteria
are considered: (1) layer distance is about 3.4\AA; (2) the shape
of $V(\theta)$ is not too complicated; (3) the difference of
maximum and minimum of $V(\theta)$ is remarkable. We find that
only (8,4) and (14,8) tube satisfy these criteria for
$0<n_1,n_2<30$ through our calculation.

The easiest way to satisfy the first condition is to put our
system into a thermal bath full of He gas whose temperature varies
with time. We will show that the outer tube will exhibit a
directional rotation.

We can write the Langevin equation \cite{Langevin} for the outer
tube $m\rho^2\ddot{\theta}=-V'(\theta)-\eta \dot{\theta}+\xi(t)$,
where $m$ and $\rho$ are respectively the mass and the radius of
the outer tube, $\eta$ is the rotating viscous coefficient
\cite{remark}, and dot and prime indicate, respectively,
differentiations with respect to time $t$ and angle $\theta$.
$\xi(t)$ is thermal noise which satisfies $\langle
\xi(t)\rangle=0$ and the fluctuation-dissipation relation
$\langle\xi(t)\xi(s)\rangle=2\eta T(t)\delta(t-s)$ \cite{Nyquist},
where $T(t)$ is temperature and the Boltzmann factor is set to 1.
Let us consider the overdamped case that the inertial term
$m\rho^2\ddot{\theta}$ is much less than thermal fluctuations and
can be neglected. Thus we arrive at $\eta\dot{\theta}=-V'(\theta)
+\xi(t)$ and the corresponding Fokker-Planck equation
\cite{Reimann,Reichl}
\begin{eqnarray}\label{fp}
\frac{\partial P(\theta,t)}{\partial
t}=\frac{\partial}{\partial\theta}\left[\frac{V'(\theta)P(\theta,t)}{\eta}\right]+\frac{T(t)}{\eta}\frac{\partial^2P(\theta,t)}{\partial\theta^2},
\end{eqnarray}
where $P(\theta,t)$ represents the probability of finding the
outer tube in angle $\theta$ at time $t$ which satisfies
$P(\theta+\pi/2,t)=P(\theta,t)$. If the period of temperature
variation is $\mathcal{T}$, we arrive at the average angular
velocity in the long-time limit \cite{Reimann}
\begin{equation}\label{current}
\langle\dot{\theta}\rangle=\lim_{t\rightarrow
\infty}\frac{1}{\mathcal{T}}\int_t^{t+\mathcal{T}}dt\int_0^{\pi/2}d\theta\left[-\frac{V'(\theta)P(\theta,t)}{\eta}\right].
\end{equation}

For example, set $T(t)=\bar{T}[1+A\sin(2\pi t/\mathcal{T})]$ with
$\bar{T}=50$ K and $|A|\ll 1$. Let $D=\eta/\bar{T}$, $t=D\tau$,
$U(\theta)=V(\theta)/\bar{T}$, $\mathcal{T}=D\mathcal{J}$ and
$\tilde{P}(\theta,\tau)=P(\theta,D\tau)$, We arrive at the
dimensionless equations of Eqs.(\ref{fp}) and (\ref{current})
\begin{eqnarray}
&&\frac{\partial \tilde{P}}{\partial
\tau}=\frac{\partial}{\partial\theta}[U'(\theta)\tilde{P}]+(1+A\sin\frac{2\pi\tau}{\mathcal{J}})\frac{\partial^2\tilde{P}}{\partial\theta^2},\label{fp2}\\
&&\langle \frac{d\theta}{d\tau}\rangle=\lim_{\tau\rightarrow
\infty}\frac{1}{\mathcal{J}}\int_{\tau}^{\tau+\mathcal{J}}d\tau\int_0^{\pi/2}d\theta\left[-U'(\theta)\tilde{P}\right].\label{cu2}
\end{eqnarray}

\begin{figure}[htp!]
\includegraphics[width=7cm]{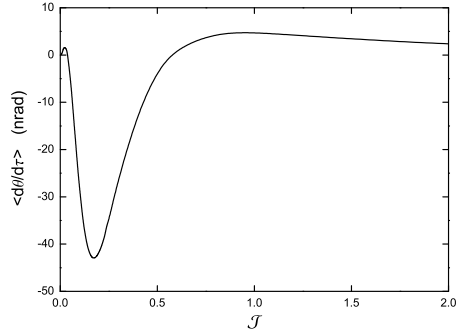}
\caption{\label{fig4}The average dimensionless angular velocity
$\langle d\theta/d\tau\rangle$ of the outer tube rotating around
the inner tube in thermal bath whose temperature changes with the
dimensionless period $\mathcal{J}$. The minus sign means the
rotation around $z$-axis is the left-handed.}
\end{figure}

Let $A=0.01$, we can numerically solve Eq.(\ref{fp2}) and
calculate Eq.(\ref{cu2}) with different $\mathcal{J}$. The solid
line in Fig.\ref{fig4} shows the relation between the average
dimensionless angular velocity
$\langle\frac{d\theta}{d\tau}\rangle$ and the dimensionless period
$\mathcal{J}$ of temperature variation. We find that
$\langle\frac{d\theta}{d\tau}\rangle\simeq 0$ for very small and
large $\mathcal{J}$, and $\langle\frac{d\theta}{d\tau}\rangle\neq
0$ for middle $\mathcal{J}$, which implies that the outer tube has
a evident directional rotation in this period range. Thus we have
constructed a temperature ratchet.

For He gas at temperature $\bar{T}=50$ K, we can calculate
$\eta=861$ Kns from its value at 273 K \cite{remark,itav} and
$D=17.2$ ns. Based on these data, we obtain
$\langle\frac{d\theta}{d\tau}\rangle=-43$ nrad when
$\mathcal{J}=0.17$, i.e.
$\langle\dot{\theta}\rangle=\frac{1}{D}\langle\frac{d\theta}{d\tau}\rangle=-2.5$
nrad/ns when $\mathcal{T}=2.9$ ns. Here the minus sign of average
angular velocity means that the rotation of the outer tube around
$z$-axis is left-handed. We notice that
$\langle\dot{\theta}\rangle=-2.5$ nrad/ns is a remarkable value
(-2.5 rad/s) that is easy to be observed from experiment. If
consider the inertial effect of the outer tube, this value may be
changed. But we believe that it is still an observable quantity.

Furthermore, from Fig. \ref{fig4} we observe that the sign of
$\langle\frac{d\theta}{d\tau}\rangle$ changes from $``+"$ to
$``-"$ and back to $``+"$ with $\mathcal{J}$ increasing, which
suggests that the outer tube's rotation around the inner tube is
the right-handed for small and large period $\mathcal{J}$, and the
left-handed for middle period $\mathcal{J}$. Through detailed
analysis, we find that $\langle\frac{d\theta}{d\tau}\rangle$ is
proportional to $\mathcal{J}^3$ for very small period
$\mathcal{J}$ and proportional to $\mathcal{J}^{-2}$ for period
$\mathcal{J}$ large enough, which agrees with the results of
asymptotic analysis \cite{Reimann,remark3}.

To summarize, we have theoretically constructed a temperature
ratchet using a DWNT, but some practical difficulties are
remained: How to synthesize the proper DWNT? How to forbid the
axial translation of outer tube? If these obstacles are ruled out,
molecular motor of DWNT must be reality.

We are grateful to H. J. Zhou for reading the manuscript and
giving much constructive advice. We thank to R. An for
mathematical discussion.

\newpage

\begin{thebibliography}{}
\bibitem{remark1} If we consider several unit cells, the final
results of this paper is unchanged.
\bibitem{remark} The rotating viscous coefficient $\eta$ is not the
viscous coeffient of gas $\eta_c$ in the common sense. Here their
relationship is $\eta=2\pi \rho^2|\mathbf{T}|\eta_c$. Generally
speaking, $\eta$ depends on the temperature (see ref.\cite{itav}),
and here temperature varies with time. Therefore $\eta$ depends on
time. But if the range of temperature variation is very small, we
can still regard $\eta$ as a constant.
\bibitem{remark3} In terms of \cite{Reimann},
$\langle\frac{d\theta}{d\tau}\rangle$ decreases proportional to
$\mathcal{J}^{-2}$ for large $\mathcal{J}$ in the case that $T(t)$
is symmetric under time inversion, and
$\langle\frac{d\theta}{d\tau}\rangle=\tilde{B}\mathcal{J}^2+\mathcal{O}(\mathcal{J}^3)$
for small $\mathcal{J}$. Here we obtain $\tilde{B}=0$ in the case
of $T(t)=\bar{T}[1+A\sin(2\pi t/\mathcal{T})]$. Thus
$\langle\frac{d\theta}{d\tau}\rangle\sim\mathcal{J}^3$ for  small
$\mathcal{J}$.
\end{thebibliography}
\end{document}